\documentclass[prl,aps,twocolumn,reprint,amsmath,amssymb,showpacs,superscriptaddress]{revtex4-1}

\usepackage{verbatim}
\usepackage{bm}
\usepackage{graphicx}
\usepackage{epsfig}
\usepackage{amsfonts}
\usepackage{amsmath}
\usepackage{amssymb}
\usepackage{subfigure}

\usepackage{makecell}
\usepackage{array}
\newcommand{\ra}[1]{\renewcommand{\arraystretch}{#1}} 
\newcolumntype{"}{@{\hskip\tabcolsep\vrule width 1pt\hskip\tabcolsep}}

\usepackage{color}

\def\cbl{\color{blue}}

\usepackage[colorlinks,bookmarks=false,citecolor=blue,linkcolor=blue,urlcolor=blue]{hyperref}

\newcommand{\be}{\begin{equation}}
\newcommand{\ee}{\end{equation}}
\newcommand{\bea}{\begin{eqnarray}}
\newcommand{\eea}{\end{eqnarray}}
\newcommand{\fig}[1]{Fig.~\ref{#1}} % fig. label

\def\vec{\mathbf}
\def\mc{\mathcal}

\allowdisplaybreaks

\usepackage{ulem}
\usepackage{times}
\begin{document}

\title{Quantum dimer model for the spin-1/2 kagome Z$_2$ spin liquid}

\author{Ioannis Rousochatzakis}
\affiliation{Institute for Theoretical Solid State Physics, IFW Dresden, D-01069 Dresden, Germany}
\affiliation{Max Planck Institut f$\ddot{u}r$ Physik Komplexer Systeme, N\"othnitzer Str. 38, 01187 Dresden, Germany}
\author{Yuan Wan}\affiliation{Department of Physics and Astronomy, The Johns Hopkins University, Baltimore, Maryland 21218}
\author{Oleg Tchernyshyov}\affiliation{Department of Physics and Astronomy, The Johns Hopkins University, Baltimore, Maryland 21218}
\author{Fr\'ed\'eric Mila}\affiliation{Institute of Theoretical Physics, \'Ecole Polytechnique F\'ed\'erale de Lausanne, CH-1015 Lausanne, Switzerland}

\date{\today}

\begin{abstract}
We revisit the description of the low-energy singlet sector of the spin-1/2 Heisenberg antiferromagnet on kagome in terms of an effective quantum dimer model. With the help of exact diagonalizations of appropriate finite-size clusters, we show that the embedding of a given process in its kagome environment leads to dramatic modifications of the amplitudes of the elementary loop processes, an effect not accessible to the standard approach based on the truncation of the Hamiltonian to the nearest-neighbour valence-bond basis. The resulting parameters are consistent with a Z$_2$ spin liquid rather than with a valence-bond crystal, in agreement with the last density matrix renormalization group results.
\end{abstract}

\maketitle

{\it Introduction} ---  
The spin $S\!=\!\frac{1}{2}$ kagome antiferromagnet (AFM) described by the Heisenberg model 
\bea\label{eqn:Ham}
\mc{H}=J\sum_{\langle ij\rangle} \vec{S}_i\cdot\vec{S}_j 
\eea
(where $\langle ij\rangle$ are nearest neighbor sites and $J\!=\!1$ in the following), has been established as one of the main candidates for realizing the long-sought quantum spin liquid~\cite{Anderson73, FazekasAnderson74,Anderson87,LiangDoucotAnderson88,Ramirez94}, which has topological properties and fractionalized excitations (spinons)~\cite{Sachdev92,Misguich02,Misguich03,Balents}. The recent discovery~\cite{Shores05} of the herbertsmithite ZnCu$_3$(OH)$_6$Cl$_2$ has motivated a new burst of experimental and theoretical activity, as this layered compound is a nearly perfect realization of (\ref{eqn:Ham}) and shows signatures of spin liquid behavior~\cite{Han12}.   

On the theory front, there are several proposals for the nature of the ground state (GS), ranging from gapless critical phases~\cite{Ran,Hermele,Ma,Iqbal}, to gapped resonating valence bond crystals (VBC)~\cite{MarstonZeng91,NikolicSenthil03,SinghHuse07,Mambrini2010a,Mambrini2010b,Syromyatnikov,Budnik,Vidal2010,PoilblancMisguich2011}, chiral~\cite{Messio2012,Capponi2012}, or topological spin liquids~\cite{YanHuseWhite,Shollwock}. In principle, such magnetically disordered phases can be described by short-range valence bonds~\cite{Anderson73,FazekasAnderson74,Anderson87,LiangDoucotAnderson88,Sandvik05}, and in fact there are good reasons to believe that the qualitative physics is captured by the nearest-neighbor valence bond (NNVB) basis. This effort goes back to Zeng and Elser~\cite{ZengElser95}, who showed that the NNVB  dynamics is governed by several tunneling processes, casted in a quantum dimer model (QDM) form~\cite{Elser1989,MambriniMila2000,Misguich03,Mambrini2010a,Mambrini2010b}. Whether this description leads to a valence-bond crystal or to a disordered phase is still open however. Indeed, recent investigations of the unconstrained QDM by Poilblanc {\it et al.}~\cite{Mambrini2010a,Mambrini2010b} and by L\"auchli~\cite{AndreasQDM} have shown that the GS is extremely sensitive to the competition between the elementary loop processes, and the answer relies on the precise determination of the corresponding QDM parameters.

\begin{figure}[!b]
\includegraphics[width=0.49\textwidth,clip]{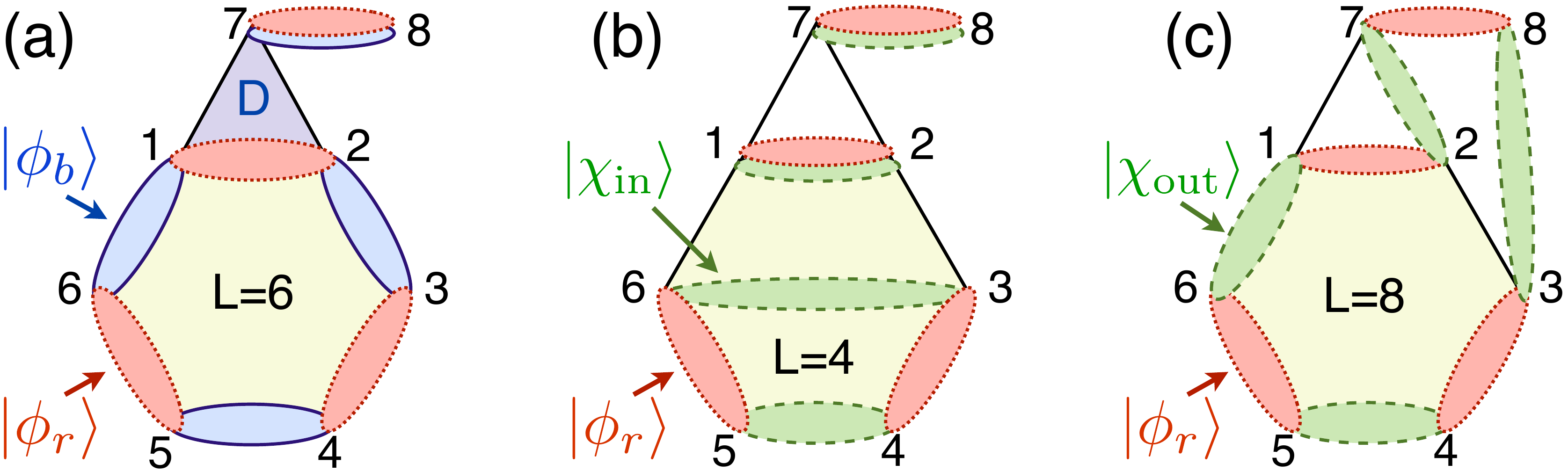}
\caption{(color online) (a) Two NNVB states on the kagome, $|\phi_b\rangle$ and $|\phi_r\rangle$, that are everywhere the same except around a hexagon (ovals denote spin singlets). (b)-(c) The VB coverings $|\chi_{\text{in}}\rangle$ and $|\chi_{\text{out}}\rangle$ (dashed green ovals) contain long-range singlets that can be virtually excited around the defect triangle `D' of $|\phi_{b}\rangle$.}\label{fig:1}
\end{figure}

\begin{figure*}[!t]
\includegraphics[width=0.999\textwidth,clip]{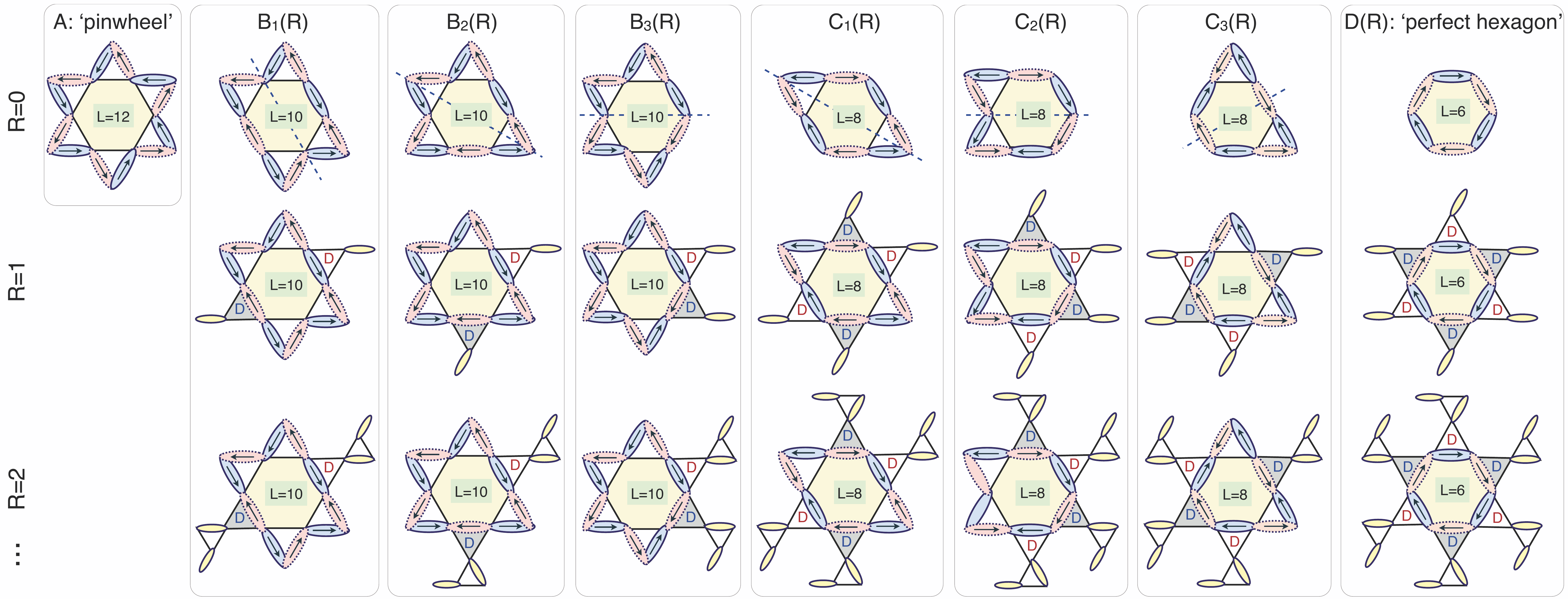}
\caption{(color online) Finite clusters used to extract the dominant tunneling parameters $t_{\text{ED}}$ on the kagome. Ovals denote singlets and arrows indicate our singlet orientation~\cite{SingletOrientation} convention (clockwise in all hexagons~\cite{Misguich02}); dashed (blue) lines are reflection symmetries. $L$ denotes the length of the loop formed by the transition graph of the two NNVB states involved in the tunneling process.}\label{fig:FiniteClusters} 
\end{figure*}

Here, we develop a new approach to determine the effective QDM parameters. This approach is based on exact diagonalizations (ED) of appropriately chosen finite clusters, and thus captures the effect of longer-range singlets that are virtually excited around triangles without a valence bond, or `defect triangles' (see Fig.~\ref{fig:1}). This method thereby captures the dependence of the QDM amplitudes on the environment (i.e. the particular VB configuration away from the central hexagon involved in the elementary loop process), by contrast to the standard approach based on the truncation of the Hamiltonian to the NNVB basis, which is shown to lead to the same parameters regardless of the environment. This effect turns out to have dramatic consequences: the QDM parameters derived in this way are consistent with a Z$_2$ spin liquid rather than a valence-bond crystal. 

Let us start by some general remarks about the QDM description of the low-energy singlet sector of a given spin-1/2 Heisenberg model. The NNVB states are in most cases linearly independent~\cite{Seidel1,Seidel2}, but the dimension of the subspace they span is much smaller than the singlet sector. Provided the low-energy states of the Heisenberg model are not orthogonal to these states (and there are no reasons for them to be), it should nevertheless be possible in principle to write down an effective Hamiltonian in this basis that reproduces the low-energy spectrum. This Hamiltonian will however in general be very complicated, with matrix elements between all states. Such a description is only useful if the resulting Hamiltonian approximately takes the form of the sum of simple local processes, as postulated by Rokhsar and Kivelson~\cite{RokhsarKivelson}, or in other words if the tunneling amplitude between dimer loops is more or less independent of the environment. So any attempt at deriving an effective QDM should address two questions: i) Can the effective model be approximated by a sum of local processes? ii) If yes, what is the value of the most important tunneling amplitudes?

The standard approach, pioneered by Rokhsar and Kivelson~\cite{RokhsarKivelson,ZengElser95,Misguich02,Mambrini2010a}, consists in truncating the Hamiltonian matrix in the NNRVB basis, and to orthogonalize the basis, leading to the effective quantum dimer Hamiltonian $\mc{H}_{\text{NNVB}}\!=\!(\mc{O}^{-\frac{1}{2}}\mc{H}\mc{O}^{-\frac{1}{2}})_{\text{NNVB}}$, where $\mc{O}$ is the overlap matrix. Assuming that the effective Hamiltonian can be written as a sum of local processes that consist in tunneling between two dimer coverings $|\phi_b\rangle$ and $|\phi_r\rangle$ (see first row of Fig.~\ref{fig:FiniteClusters} for a sketch of the most important processes), the QDM parameters can in fact be simply determined  from the overlap $\langle \phi_b|\phi_r\rangle\!\equiv\!\omega$ and the matrix elements $\langle \phi_b|\mc{H}|\phi_b\rangle\!=\!\langle \phi_r|\mc{H}|\phi_r\rangle\!\equiv\!E_0$ and $\langle \phi_b|\mc{H}|\phi_r\rangle\!\equiv\!v$, which can be found from the transition graph following standard rules~\cite{Sutherland88, RokhsarKivelson,ZengElser95,MambriniMila2000, Misguich02,Mambrini2010a}. Indeed, the simplicity of the problem allows for a direct evaluation of $\mc{O}_{\text{NNVB}}^{-\frac{1}{2}}$, giving 
\be\label{eqn:HNNVB}
\mc{H}_{\text{NNVB}}= \left(
\begin{array}{cc}
 E_0+V_{\text{NNVB}} & t_{\text{NNVB}} \\
t_{\text{NNVB}} &   E_0+V_{\text{NNVB}}
\end{array}
\right)~,
\ee
with 
\be\label{eq:tV}
t_{\text{NNVB}}=\frac{v-E_0\omega}{1-\omega^2},~~\text{and}~~V_{\text{NNVB}}=- t_{\text{NNVB}}~\omega~.
\ee
These expressions are identical with Eq.~(40) of \cite{Mambrini2010a}, up to different conventions for the dimer orientation and for the Hamiltonian~\cite{SM}. This means that for elementary processes the infinite linked-cluster expansion of \cite{Mambrini2010a} is equivalent to truncating to the 2$\times$2 NNVB basis of the corresponding finite cluster. 

In this approach, the values of $t_{\text{NNVB}}$ and $V_{\text{NNVB}}$ do not depend at all on the embedding: Indeed, for a given process, clusters corresponding to different rows of Fig.~\ref{fig:FiniteClusters} have different $E_0$ and $v$, but the same $\omega$ and $v\!-\!E_0\omega$. So, formulated in this way, the truncation approach leads to the intriguing conclusion that there is no embedding dependence whatsoever. 

\begin{figure*}[!t]
\includegraphics[width=0.96\textwidth,clip]{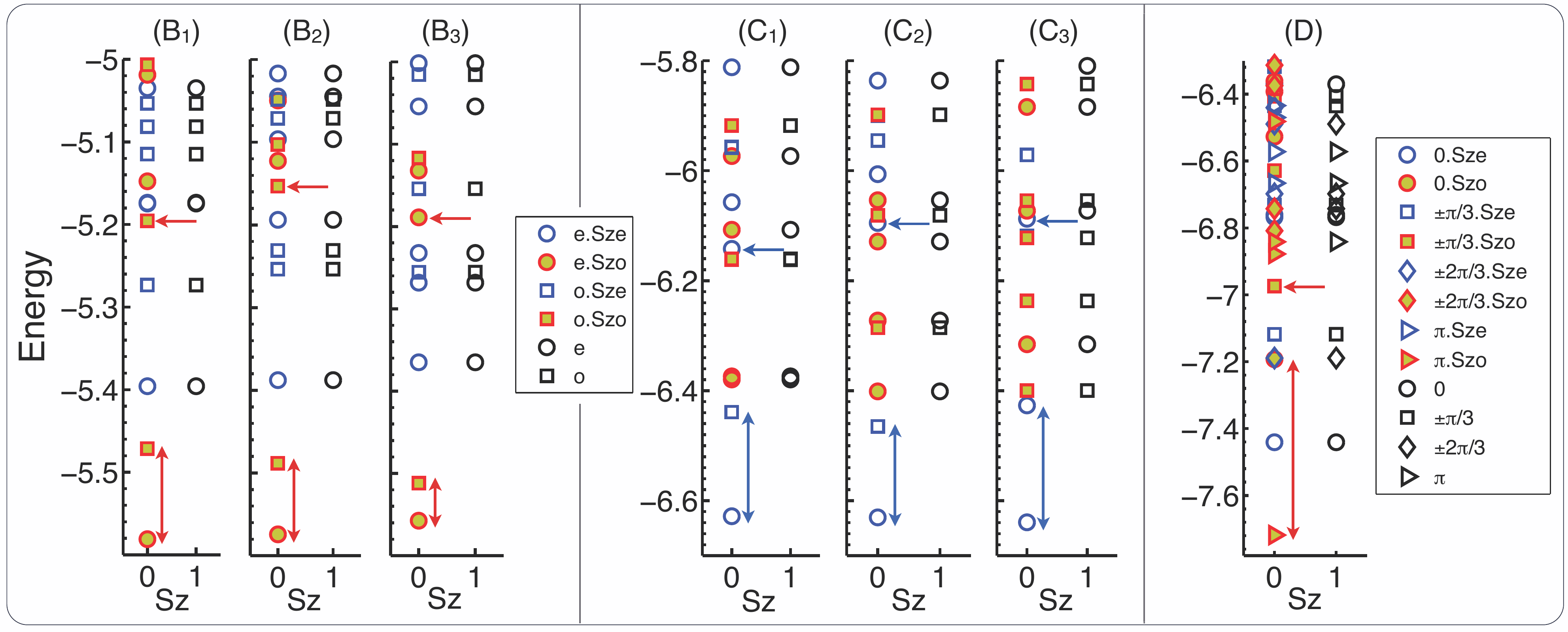}
\caption{(color online) Low-energy spectra of the `$R\!=\!1$' clusters B$_1$-D of \fig{fig:FiniteClusters}  in the magnetization sectors $S_z\!=\!0$ and $1$. Legends: parity under reflection (for B$_1$-C$_3$), where ``e'' and ``o'' denote even and odd parity respectively or, angular momenta under six-fold rotations (for D). Spin-inversion parity (for $S_z\!=\!0$) is labeled by ``Sze'' and ``Szo''. The two NNVB low-lying singlets are designated by double vertical arrows and the third singlet by horizontal arrows. 
\label{fig:Spectra}} 
\end{figure*}

If this conclusion was correct, it would mean that the low-energy singlet sector of the Hamiltonian on clusters with different embeddings should be identical up to an overall constant. To challenge this conclusion, a natural thing to do is to enlarge the truncation basis to include the {\it intermediate} longer-range singlets that are generated by applying the Hamiltonian, like $|\chi_{\text{in}}\rangle$ for a single hexagon, or  $|\chi_{\text{in}}\rangle$ and $|\chi_{\text{out}}\rangle$  for the 8-site cluster of Fig.~\ref{fig:1}. Such a basis, denoted by `NNVB+', leads to an effective Hamiltonian $\mc{H}_{\text{NNVB+}}\!=\!(\mc{O}^{-1/2}\mc{H}\mc{O}^{-1/2})_{\text{NNVB+}}$ that includes the renormalization from the intermediate long-range singlets and thus the dependence on the environment. It turns out that the low-energy sectors of $\mc{H}_{\text{NNVB,+}}$ for these two clusters are indeed different from each other as well as from that of the fully truncated Hamiltonian $\mc{H}_{\text{NNVB}}$~\cite{SM}. In particular, the splitting between the low-lying singlets evolves from $1.2$ for $\mc{H}_{\text{NNVB}}$, to $\frac{\sqrt{13}-1}{2}\!=\!1.30277$ for $\mc{H}_{\text{NNVB,+}}$ of the single hexagon, and finally to $1.15697$ for $\mc{H}_{\text{NNVB,+}}$ of the cluster of Fig.~\ref{fig:1}. The corresponding reduction for the minimal embedding of the hexagon in the kagome (second row and last column of Fig.~\ref{fig:FiniteClusters}) is even larger, on account of the many more intermediate states $|\chi_{\text{out}}\rangle$ involved (see below).
This example shows that there has to be a significant embedding dependence of the tunneling amplitude since it controls the splitting. The rest of the paper is devoted to a systematic investigation of this dependence.

To this end, we imagine embedding the minimal clusters shown in the first row of Fig.~\ref{fig:FiniteClusters} back on the kagome in such a way that the resulting clusters can only accommodate the two dimer coverings involved in each process. In choosing the clusters that allow to capture the most important contributions, the spinon picture of the VB dynamics of \cite{Hao10} offers a clear insight. According to this picture, a defect triangle is a bound state of two antikink spinons, and the virtual VB dynamics is due to the motion of such spinons. Following \cite{Hao10}, antikinks can only escape our clusters through the dangling (yellow) bonds of the second row of Fig.~\ref{fig:FiniteClusters}. So to capture the most important fluctuations and to simulate the kagome as close as possible, we proceed by attaching open sawtooth chains of length $R$ next to the defect triangles of the minimal `$R\!=\!0$' clusters. The influence of the remaining environment (most notably, the possible presence of extra defect triangles in the immediate vicinity of the central hexagon) will also be discussed below.

\begin{table*}[!t] %\ra{0.5}
\begin{ruledtabular}
\caption{Tunneling amplitude $t_{\text{ED}}$ for the most local processes on the kagome, as extracted from exact diagonalization spectra of the clusters shown in \fig{fig:FiniteClusters}. For comparison, we also give the amplitudes from the NNVB projection (second column).}\label{tab:tED}
\begin{tabular}{l|ccccccc} 
process &  NNVB &$R=0$ & $R=1$ & $R=2$  &  $R=3$  &  $R=4$  & $R=5$ \\
\hline
A & 0 & 0 & 0 & 0 & 0 & 0 & 0 \\
B$_1$ &$\frac{4}{85}=0.04705$& +0.09230 &+0.05476 & +0.04779 & +0.04511 & +0.04334 & +0.04197\\
B$_2$ &$\frac{4}{85}=0.04705$& +0.06216 &+0.04301 & +0.03957 & +0.03812 & +0.03701 & +0.03608\\
B$_3$ &$\frac{4}{85}=0.04705$& +0.01952 &+0.02272 & +0.02496 & +0.02562 & +0.02564 & +0.02544\\
C$_1$ &$-\frac{4}{21}=-0.19047$& -0.11726 &-0.09461 & -0.08246 &  -0.07028 &&\\
C$_2$ &$-\frac{4}{21}=-0.19047$& -0.11586 &-0.08252 & -0.06735 & -0.05737&&\\
C$_3$ &$-\frac{4}{21}=-0.19047$& -0.19152 &-0.10634 & -0.08410 & -0.07113&&\\
D  & $\frac{3}{5}=0.6$&$\frac{\sqrt{13}-1}{4}=0.65138$ &+0.26285 & +0.14664 &&&
\end{tabular}
\end{ruledtabular}
\end{table*}

When we enlarge the embedding along these lines, the number of intermediate longer-range singlets that need to be included in the truncation basis increases very quickly, rendering the analytical calculation of the matrix elements of $\mc{H}_{\text{NNVB+}}$ practically intractable.  
However, a completely equivalent but technically much simpler method is to use numerical ED in the full singlet basis and extract the effective QDM parameters from the lowest two singlets in the spectrum. 
The basic idea is that, when the virtual potential energy corrections $V_r$ and $V_b$ of $|\phi_r\rangle$ and $|\phi_b\rangle$, respectively, are the same, then the tunneling amplitude is half the splitting. Now, except for the cluster C$_3$, in which the reflection operation maps $|\phi_b\rangle\!\mapsto\!|\phi_b\rangle$ and $|\phi_r\rangle\!\mapsto\!|\phi_r\rangle$, in all remaining clusters the reflection (B$_1$-C$_2$) or the six-fold rotation (A, D) map $|\phi_b\rangle\!\mapsto\!\eta|\phi_r\rangle$, where $\eta\!=\!\pm1$. This means that $V_r\!=\!V_b$ and, moreover, the two eigenstates of the effective QDM, $|\pm\rangle\!=\!\frac{|\phi_b\rangle\pm|\phi_r\rangle}{2\sqrt{1\pm\omega}}$, have well defined parity $\pm\eta$. So we can extract the tunneling amplitude $t$, including its sign, 
by simply taking 
\be\label{eq:tED}
t_{\text{ED}}=\frac{E_+\!-\!E_-}{2}, 
\ee
where $E_+$ and $E_-$ are the energies corresponding to $|\pm\rangle$, which in turn can be identified in the spectrum by their parity. 

For cluster C$_3$, $|\phi_r\rangle$ and $|\phi_b\rangle$ are not related by any symmetry and thus $V_r\!\neq\!V_b$~\cite{DTDistance}, and $\frac{E_+\!-\!E_-}{2}\!=\![(\frac{V_r\!-\!V_b}{2})^2\!+\!t_{\text{ED}}^2]^{1/2}$. Thus we need an independent calculation for $V_r\!-\!V_b$ in order to extract $t_{\text{ED}}$. As we show in \cite{SM}, $|V_r\!-\!V_b|\!\approx\!0.005$, which is much smaller than the splitting $|E_+\!-\!E_-|$ found for C$_3$. So to a very good approximation, $|t_{\text{ED}}|\!\approx\!\frac{|E_+\!-\!E_-|}{2}$. The sign of $t_{\text{ED}}$ is assumed the same with that of $t_{\text{NNVB}}$, as happens for the other processes.

\begin{figure}[!t]
\includegraphics[width=0.48\textwidth]{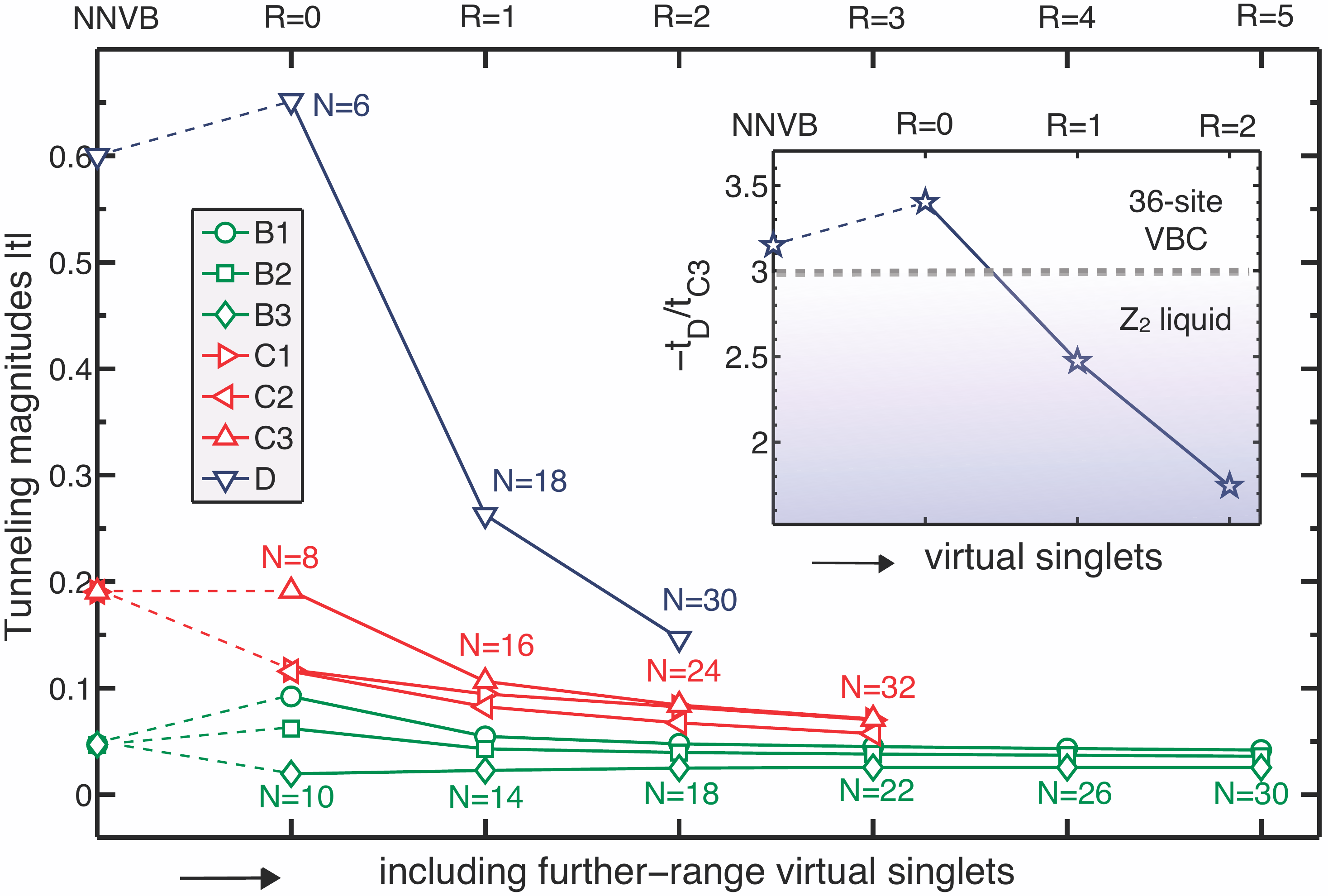}
\caption{(color online) (a) Evolution of $|t|$ as we include further-range virtual singlet fluctuations, quantified by the length $R$ of the sawtooth chains attached to the minimal clusters of the first row  of Fig.~\ref{fig:FiniteClusters}. The NNVB parameters are placed at the origin for comparison. Inset: The ratio of the two strongest processes. The horizontal dashed (gray) line separates the 36-site VBC from the $Z_2$ spin liquid phase~\cite{AndreasQDM}.}\label{fig:tvsR}
\end{figure}

Some examples of spectra are shown in Fig.~\ref{fig:Spectra} for the `$R\!=\!1$' clusters, the values of the tunneling amplitudes deduced in that way are summarized in Table~\ref{tab:tED}, and their evolution with the `size' of the embedding is plotted in Fig.~\ref{fig:tvsR}. These data, which are the main results of the present paper, contain a lot of important information.

First of all, the singlet sector of the exact spectra shown in Fig.~\ref{fig:Spectra} is consistent with the NNVB picture, as we can clearly identify two low-lying singlets (double vertical arrows) with the expected symmetry, that are separated from the third singlet (horizontal arrow) by an appreciable energy gap~\cite{LETriplets}. 
Second, the tunneling amplitude {\it does} depend on the environment, as expected. The NNVB truncation approach is in reasonable agreement only with the `$R\!=\!0$' clusters C$_3$ and D, and so it can be considered as an approximate way to calculate the tunneling amplitudes $t_{C_3}$ and $t_D$ without any embedding. However, to describe the kagome antiferromagnet, embedding should be taken into account, and it clearly matters, as demonstrated by the evolution of the parameters with $R$.

Third, unlike $t_{\text{B}_{1-3}}$ and $t_{\text{C}_{1-3}}$, which converge quickly with $R$, the amplitude $t_\text{D}(R)$ for the `perfect hexagon' process shows a remarkable drop from $0.65138$ at $R\!=\!0$ to $0.26285$ at $R\!=\!1$, and a similarly high drop to $0.14664$ at $R\!=\!2$. This shows that fluctuations have the strongest influence on the process with the largest number of defect triangles (and spinons~\cite{spinons}). Hence, the 36-site VBC favored by the `perfect hexagon' process D is clearly disfavored by virtual fluctuations (see below). 
Note that the `$R\!=\!3$' cluster of type (D) has $N\!=\!42$ sites, which is too large for an ED treatment, given the small number of available symmetries. However, the three available ED data suggest a convergence at $R\!\gtrsim\!4\!-\!5$, and it is likely that its value will converge slightly below that of $t_{\text{C}_{1-3}}$. 
The reason why convergence is slowest in (D) is related to statistical pressure (spinons are fermions on kagome~\cite{Hao10}) which tends to push spinons out of the clusters;  the more spinons the higher the pressure, and thus the longer the chains needed to accomodate the escaping spinons. 

Fourth, as we show in \cite{SM}, already for `$R\!=\!2$' clusters, the tunneling amplitude of some processes starts to depend on the remaining environment besides the sawtooth chains. This shows that a description in terms of a QDM with `local' processes can only be approximate. However, most processes still have the same amplitude at the level `$R\!=\!2$', and those which take different values have a small dispersion except for very rare configurations with an anomalously large concentration of defect triangles. So it is still possible to define an approximate description with `average' tunneling amplitudes. 

The potential terms of the QDM can be extracted in a similar way with the help of a numerical linked-cluster procedure~\cite{SM}. The resulting data show that the binding energy of two defect triangles $V_{2\text{-dt}}$ can be essentially neglected, but the binding energy $V_{3\text{-dt}}$ between three defect triangles around a hexagon (process D) is large and positive. The specific values depend again on the embedding ($0.18470$ for $R\!=\!1$ and $0.09718$ for $R\!=\!2$), but they are of the same order of magnitude as the corresponding tunneling, so they must be taken into account. Importantly, the fact that $V_{3\text{-dt}}$ is positive (in contrast to series expansions around the perturbative dimerized limit~\cite{SinghHuse07}) means that virtual fluctuations penalize  the `perfect hexagons' not only by the strong reduction in $t_D$ but also via the potential terms.

{\it Discussion} --- 
In their recent study of the unconstrained QDM model, Poilblanc {\it et al.}~\cite{Mambrini2010a, Mambrini2010b} interpolated between the Heisenberg point (in its truncated NNVB treatment) and the exactly solvable spin liquid point of Misguich {\it et al.}~\cite{Misguich02}, while L\"auchli~\cite{AndreasQDM} examined the fate of the QDM as a function of the ratio between the two strongest tunneling processes, i.e. $t_{\text{D}}$ and $t_{\text{C}}$. Both studies showed that the system is extremely close to a quantum critical point, separating the 36-site VBC state from a Z$_2$ spin liquid phase. While it is possible that the two liquid phases are adiabatically connected to each other, the path in the parameter space taken in \cite{AndreasQDM} is more directly connected to the present findings, since $t_{\text{D}}$ and $t_{\text{C}_{1-3}}$ are the strongest and also the ones that are affected the most by virtual fluctuations. According to \cite{AndreasQDM}, the GS of the model turns to a Z$_2$ spin liquid already at $-t_{\text{D}}/t_{\text{C}}\sim 3$, and including the potential terms would probably shift this boundary even higher. In any case, as shown in Fig.~\ref{fig:tvsR} (inset), the ratio $-t_{\text{D}}/t_{\text{C}_3}$ goes way below this boundary, meaning that we are deep inside the Z$_2$ spin liquid phase. It would be interesting to check explicitly whether the more precise model that includes the dependence on the remaining environment and the potential terms leads to the same phase, a plausible result since the average model is deep in the Z2 spin liquid phase. This is left for future investigation.

{\it Acknowledgments:} 
We would like to thank A. M. L\"auchli for enlightening discussions and M. Mambrini for fruitful correspondence. This work was partly supported by the Deutsche Forschungsgemeinschaft (DFG) under the Emmy-Noether program, by the Swiss National Fund, and by the U.S. Department of Energy, Office of Basic Energy Sciences, Division of Materials Sciences and Engineering under Award DE-FG02-08ER46544.

%---------------------------------------------------------------------------------
\clearpage

\appendix

\pagenumbering{roman}
\widetext

\begin{widetext}
\section{Supplemental material}

\section{A. Basic elements of the NNVB truncation method}
Let us consider an elementary tunneling process between two dimer coverings $|\phi_b\rangle$ and $|\phi_r\rangle$. For the dimer orientations, we follow the convention of Misguich {\it et al.}\cite{Misguich03S} and take the singlets to be oriented clockwise in each hexagon, see Fig.~2 in main text. Let us suppose further that the transition graph~\cite{Sutherland88S} of our elementary process involves a single non-trivial loop of length $L$, surrounding $N_{\text{hex}}$ hexagons. Then our convention gives for the overlap~\cite{Misguich03S}: 
\be\label{eq:omega1}
\omega =  \langle \phi_r|\phi_b\rangle = (-1)^{\frac{L}{2}+N_{\text{hex}}+1}~2^{1-\frac{L}{2}} ~.
\ee
For the eight most local and dominant processes (shown in Fig.~2 of the main tex), $N_{\text{hex}}=1$ and so the sign of $\omega$ is $(-1)^{\frac{L}{2}}$, see Table~\ref{tab:NNVB}.

\begin{table}[!b] \ra{1.3}
\begin{ruledtabular}
\caption{Basic NNVB parameters for each of the eight most local tunneling processes on the kagome (Fig.~2 of main text). The amplitudes $t_{\text{NNVB}}$ depend on three basic quantities, $\omega$, $E_0$, and $v$, whose calculation follows from standard rules~\cite{Sutherland88S,ZengElser95S,Misguich03S,Mambrini2010aS}. For the overlap signs we have chosen the convention that singlets are oriented clockwise in each hexagon. Note that the values of $E_0$ and $v$ depend on $R$ (here we give the values for $R=1$) but the values of $\omega$, $t_{\text{NNVB}}$ and $V_{\text{NNVB}}$ do not. The tunneling amplitudes of Ref.~[\onlinecite{Mambrini2010aS}] equal $-\frac{4}{3}t_{\text{NNVB}}$ (sixth column).}\label{tab:NNVB}
\begin{tabular}{l|cccccr} 
& $L$ & $\omega$ & $-\frac{4}{3}E_0$ & $-\frac{4}{3}v/\omega$  &  $-\frac{4}{3}t_{\text{NNVB}}$  &  $t_{\text{NNVB}}$ \\
\hline
A   &12& $2^{-5}$ & $6$ & $6$ & $0$ & 0\\
B$_1$ &10& $-2^{-4}$ & $7$ & $8$ & $\frac{\omega}{1-\omega^2}=-\frac{16}{255}$ & $+\frac{4}{85}$\\
B$_2$ &10& $-2^{-4}$ & $7$ & $8$ & $\frac{\omega}{1-\omega^2}=-\frac{16}{255}$ & $+\frac{4}{85}$\\
B$_3$ &10& $-2^{-4}$ & $7$ & $8$ & $\frac{\omega}{1-\omega^2}=-\frac{16}{255}$ & $+\frac{4}{85}$\\
C$_1$ &8& $2^{-3}$ & $8$ & $10$ & $2\frac{\omega}{1-\omega^2}=\frac{16}{63}$ & $-\frac{4}{21}$\\
C$_2$ &8& $2^{-3}$ & $8$ & $10$ & $2\frac{\omega}{1-\omega^2}=\frac{16}{63}$ & $-\frac{4}{21}$\\
C$_3$ &8& $2^{-3}$ & $8$ & $10$ & $2\frac{\omega}{1-\omega^2}=\frac{16}{63}$ & $-\frac{4}{21}$\\
D   &6& $-2^{-2}$ & $9$ & $12$ & $3\frac{\omega}{1-\omega^2}=-\frac{4}{5}$ & $+\frac{3}{5}$ 
\end{tabular}
\end{ruledtabular}
\end{table}

Let us now compare with the convention of Schwandt {\it et al.}~\cite{Mambrini2010aS} Here the authors choose a fermionic representation of the singlet wavefunction, which relies on a particular site ordering choice. For processes involving a single non-trivial loop, this convention leads to~\cite{Mambrini2010aS}:
\be\label{eq:omega2}
\omega = (-1)^{\frac{L}{2}+1}~2^{1-\frac{L}{2}}~, 
\ee
which differs from Eq.~(\ref{eq:omega1}) by a factor $(-1)^{N_{\text{hex}}}$. So the two conventions deliver the opposite overlap signs for processes involving an odd number of hexagons, like the eight most local processes (Fig.~2 of the main text). 

To see what the above conventions give for the signs of $t_{\text{NNVB}}$ and $V_{\text{NNVB}}$ we rewrite Eq.~(3) of the main text as: 
\be\label{eq:tV}
t_{\text{NNVB}} = + h \frac{\omega}{1-\omega^2},~~V_{\text{NNVB}}=-t_{\text{NNVB}} ~\omega
\ee
with $h=v/\omega-E_0$. (This quantity corresponds to Eq.~(12) of Schwandt {\it et al}~\cite{Mambrini2010aS}, modulo a factor of 3/4 related to the redefinition of the Hamiltonian; see passage below Eq.~(10) in [\onlinecite{Mambrini2010aS}]. So Eq.~(\ref{eq:tV}) is consistent with Eq.~(40) of [\onlinecite{Mambrini2010aS}]). 
It follows that, for the eight most local processes, the two conventions lead to opposite signs for $\omega$ and $t_{\text{NNVB}}$ but the same signs for $V_{\text{NNVB}}$.

%%%%%%%%%%%%%%%%%%%%%%%%%%%%%%

\section{B. Truncation approach for the `perfect hexagon' tunneling process}
Here we present three different levels of the truncation approach for the `perfect hexagon' process D. 
First we summarize the results from the fully truncated basis of the two NNVB states  
\be
\parbox{2.1in}{\epsfig{file=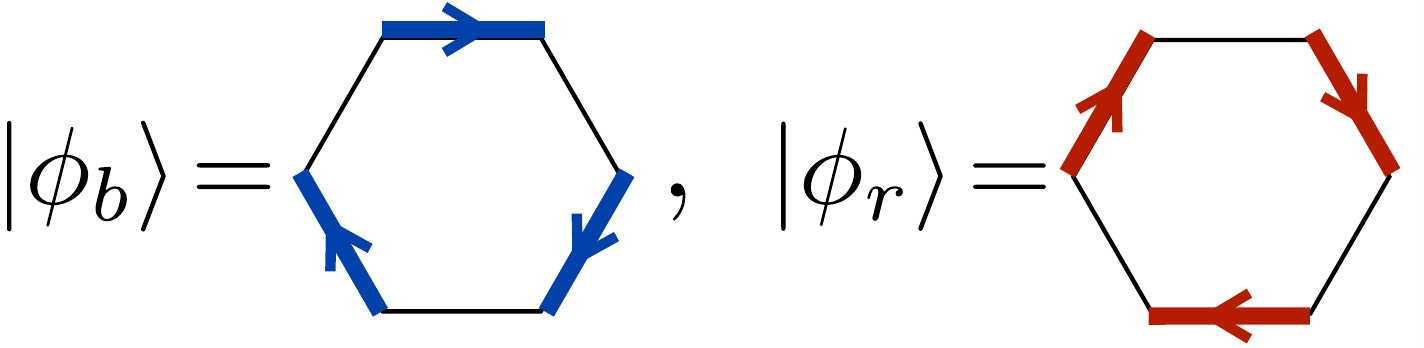,width=2.1in,clip=}}~.
\ee
In this basis we find
\be
\mc{O}\!=\!\left(\!\!
\begin{array}{rr}
1 & -1/4 \\
-1/4 & 1
\end{array}
\!\!\right),~~~
\mc{H}\!=\!\left(\!\!
\begin{array}{rr}
-9/4 & +9/8 \\
+9/8 & -9/4
\end{array}
\!\!\right),~~~
\mc{H}_{\text{NNVB}}\!=\!\mc{O}^{-1/2}\mc{H}\mc{O}^{-1/2}\!=\!\left(\!\!
\begin{array}{rr}
-21/10 & +3/5 \\
+3/5 & -21/10
\end{array}
\!\!\right)~. \nonumber
\ee
Diagonalizing $\mc{H}_{\text{NNVB}}$ gives the eigenvalues 
\be\label{eq:ENNVB}
\mc{E}_{\text{NNVB}}=\{-\frac{27}{10},-\frac{3}{2}\}~.
\ee
A full diagonalization in the entire singlet sector of the problem, which contains five singlets, gives the eigenvalues 
\be\label{eq:EfullH}
\mc{E}_{\text{full}}=\{-\frac{\sqrt{13}}{2}-1,-\frac{3}{2},-\frac{1}{2},-\frac{1}{2},\frac{\sqrt{13}}{2}-1\}~.
\ee
So, the truncation into the NNVB basis does not reproduce the exact result for the ground state energy, but only for the first excited singlet (with energy $-3/2$), which corresponds to the symmetric combination $|\phi_b\rangle+|\phi_b\rangle$ (by symmetry, this state does not couple to any other state).

Next, we enlarge the truncation basis so that we include the intermediate longer-range singlets that are generated by applying the Hamiltonian onto the NNVB basis. For a single hexagon, the extended basis `NNVB+'  involves three states in total:
\be
\parbox{4.5in}{\epsfig{file=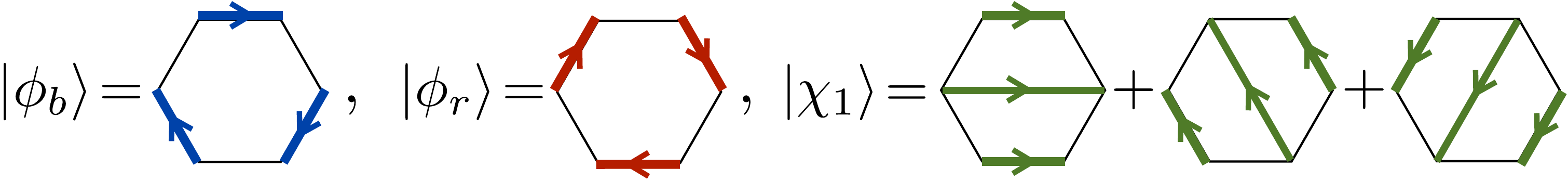,width=4.5in,clip=}}~.
\ee
i.e., there is only one intermediate longer-range state, $|\chi_1\rangle$. In this basis we find:
\be
\mc{O}\!=\!\left(\!\!
\begin{array}{rrr}
1 & -1/4 & -3/2 \\
-1/4 & 1 & +3/2\\
-3/2 & +3/2 & 9/2
\end{array}
\!\!\right),~~
%%%
\mc{H}\!=\!\left(\!\!
\begin{array}{rrr}
-9/4 & +9/8 & +9/2 \\
+9/8 & -9/4 & -9/2\\
+9/2 & -9/2 & -45/4
\end{array}
\!\!\right)~.\nonumber
\ee
Diagonalizing the effective QDM Hamiltonian $\mc{H}_{\text{NNVB+}}\!=\!\mc{O}^{-1/2}\mc{H}\mc{O}^{-1/2}$ gives the eigenvalues 
\be
\mc{E}_{\text{NNVB+}}=\{-\frac{\sqrt{13}}{2}-1,-\frac{3}{2},\frac{\sqrt{13}}{2}-1\}~.
\ee
Comparing to Eq.~(\ref{eq:EfullH}), we see that the truncation into the `NNVB+' basis reproduces the exact results for all three singlets involved in the tunneling physics.

To include more intermediate states we now turn to the cluster of Fig.~1 of the main text. Here, the `NNVB+' basis contains seven singlets, namely the two NNVB states plus five intermediate longer-rangle singlets:
\be
\parbox{5.5in}{\epsfig{file=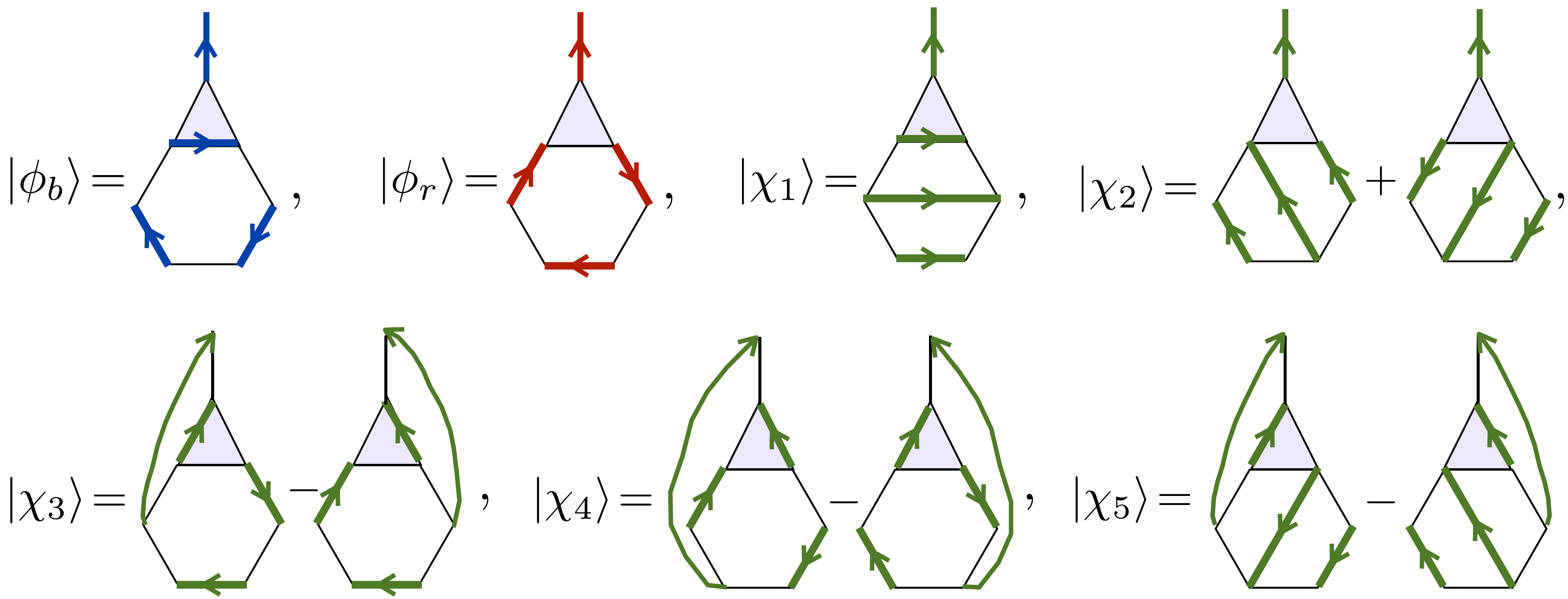,width=5.5in,clip=}}
\ee
In this basis we find:
\be
\mc{O}\!=\!\left(\!\!
\begin{array}{rrrrrrr}
1 & -1/4 & -1/2 &-1&1/4& 1/2 & -1/2\\
-1/4 & 1 & 1/2 &1&-1&-1/2&1/2\\
-1/2 & 1/2 & 1 &1/2&-1/2&-1/4&1/4\\
-1&1&1/2&5/2&-1&-5/4&5/4\\
1/4&-1&-1/2&-1&5/2&5/4&-5/4\\
1/2&-1/2&-1/4&-5/4&5/4&7/4&-1\\
-1/2&1/2&1/4&5/4&-5/4&-1&7/4\\
\end{array}
\!\!\right)\!\!,~
%%%
\mc{H}\!=\!\left(\!\!
\begin{array}{rrrrrrr}
-3&21/16&15/8&15/4&-21/16&-15/8&15/8\\
21/16&-3&-15/8&-15/4&15/4&9/4&-9/4\\
15/8&-15/8&-9/4&-21/8&15/8&21/16&-21/16\\
15/4&-15/4&-21/8&-57/8&9/2&69/16&-69/16\\
-21/16&15/4&15/8&9/2&-51/8&-63/16&63/16\\
-15/8&9/4&21/16&69/16&-63/16&-63/16&3\\
15/8&-9/4&-21/16&-69/16&63/16&3&-51/16\\
\end{array}
\!\!\right).\nonumber
\ee
Diagonalizing the effective QDM Hamiltonian $\mc{H}_{\text{NNVB+}}\!=\!\mc{O}^{-1/2}\mc{H}\mc{O}^{-1/2}$ gives the eigenvalues 
\be\label{eq:ENNVBp8site}
\mc{E}_{\text{NNVB+}}=\{ -3.62372, -2.46675, -1.56287, -0.973615,-0.846768,-0.2908, 1.01452\}~.\nonumber
\ee
A full diagonalization in the entire singlet sector of the problem, which now contains fourteen singlets, gives the eigenvalues 
\bea\label{eq:Efull8site}
\mc{E}_{\text{full}}=\{&&
{\cbl-3.62372}, -2.47474, {\cbl-2.46675}, -1.70996, {\cbl-1.56287}, -1.33712,{\cbl-0.973615}, -0.871334, \nonumber\\
&&
{\cbl-0.846768}, {\cbl-0.2908}, -0.0252551, 0.360045, {\cbl1.01452}, 1.30836\}~.\nonumber
\eea
Again the truncation into the `NNVB+' basis reproduces the exact energies for the seven singlets involved in the tunneling physics. Note that the `intruder' state with energy $E\!=\!-2.47474$ has even parity under the reflection symmetry of the cluster, while the first ($E\!=\!-3.62372$) and the third ($E\!=\!-2.46675$) have even and odd parity, respectively, as expected from the symmetry of the two tunnel-split NNVB states.

%%%%%%%%%%%%%%%%%%%%%%%%%%%%%%
\section{C. Influence of extra defect triangles in the tunneling amplitudes}
In the main text we considered clusters with open delta-chains that are attached at the positions of the defect triangles of the `$R\!=\!1$' clusters. This procedure neglects the possible presence of extra defect triangles nearby. The closest position where these extra defect triangles may sit are right next to the elementary loop. 
Let us take, for example, the cluster C$_1$ at the `$R\!=\!1$' level. There are three possibilities: no extra defect triangle (C$_1^{(0)}$), one extra defect triangle (C$_1^{(1)}$), or two extra defect triangles (C$_1^{(2)}$): 
\be
\parbox{5.in}{\epsfig{file=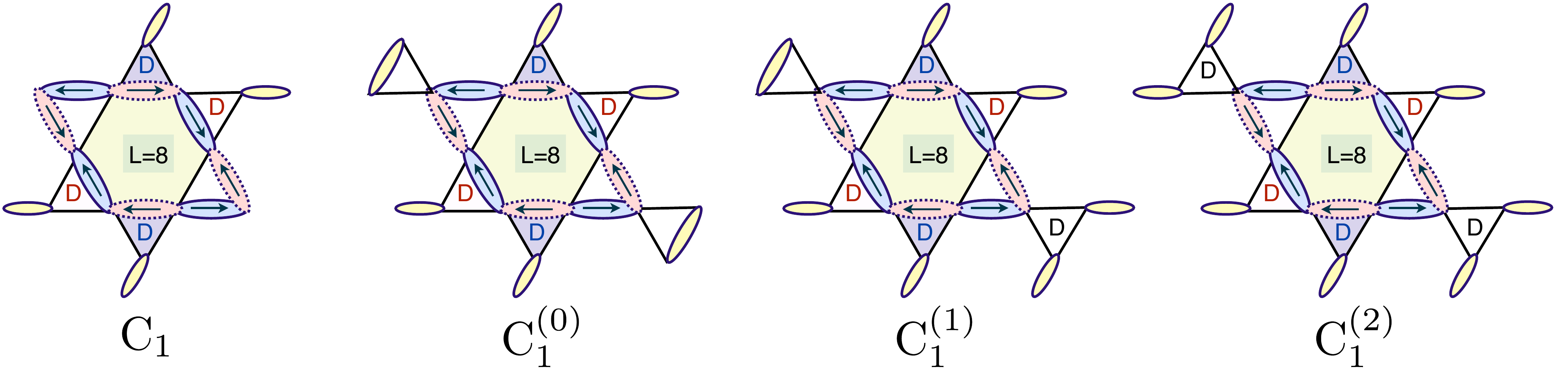,width=5.in,clip=}}
\ee
Similarly, there are three possibilities for the processes C$_2$ and C$_3$. The tunneling amplitudes extracted for C$_{1-3}$ are the same with the ones extracted from C$_{1-3}^{(0)}$, respectively. This happens because the extra triangles that are present in C$_{1-3}^{(0)}$ (compared to C$_{1-3}$) have a valence bond which, not only minimizes exactly the extra terms of the Hamiltonian, but also remains intact no matter how many times we apply the Hamiltonian. So the tunneling amplitudes discussed in the main text correspond to the situation without any extra defect triangles.

The tunneling amplitude can again be extracted from the exact spectra of these clusters, as explained in the main text. We find:  
$t(\text{C}_1^{(0)})\!=\!-0.09461$, 
$t(\text{C}_1^{(1)})\!=\!-0.06708$, and 
$t(\text{C}_1^{(2)})\!=\!-0.04208$; 
$t(\text{C}_2^{(0)})\!=\!-0.08252$, 
$t(\text{C}_2^{(1)})\!=\!-0.05831$, and 
$t(\text{C}_2^{(2)})\!=\!-0.03580$;
$t(\text{C}_3^{(0)})\!=\!-0.10634$, 
$t(\text{C}_3^{(1)})\!=\!-0.08739$, and 
$t(\text{C}_3^{(2)})\!=\!-0.07752$.  
These numbers will be further modified by enlarging the length $R$ as we saw in the main text, so they only give a representative picture for the influence of the extra defect triangles. We see that the extra defect triangles reduce the magnitude of $t$ by an amount that can be comparable to the one from the $R\!>\!1$ portion of the attached chains. 

Now, the tunneling processes C$_1$-C$_3$ contain already two defect triangles and so having one or two extra defect triangles nearby amounts to a high density of defect triangles, since on average we should encounter one defect triangle every four triangles (the total percentage of defect triangles is 1/4 in every NNVB state). So these cases can be considered rare. Moreover, the fact that the environments with no extra defect triangles gives the largest $|t|$ means that such environments will be energetically preferred.

The processes B$_1$-B$_3$ involve one defect triangle, so the presence of extra defect triangles nearby is more likely here. We have seen that these processes are much weaker than C$_1$-C$_3$ and D, and the fluctuations along the delta chains and the extra defect triangles will reduce it even further, meaning that these processes do not play much role anyway. Finally, the `perfect hexagon' process D contains already the maximum number of defect triangles in the central hexagon and the possibility of having extra defect triangles in the immediate vicinity of the hexagon is therefore absent.

%%%%%%%%%%%%%%%%%%%%%%%%%%%%%%
\section{D. Potential terms}
\subsubsection{D1. Potential terms involving a single defect triangle}
The potential terms $V_{1\text{-dt}}$ that come from processes around a single defect triangle give rise to a global energy shift of the spectrum, because the total number of defect triangles is the same for all NNVB states. Still, the value of $V_{1\text{-dt}}$ is important if we want to extract the irreducible contribution to the binding energy between two or more defect triangles. The minimal cluster involving a single defect triangle is 
\be
X_0 = \!\!\parbox{0.7in}{\epsfig{file=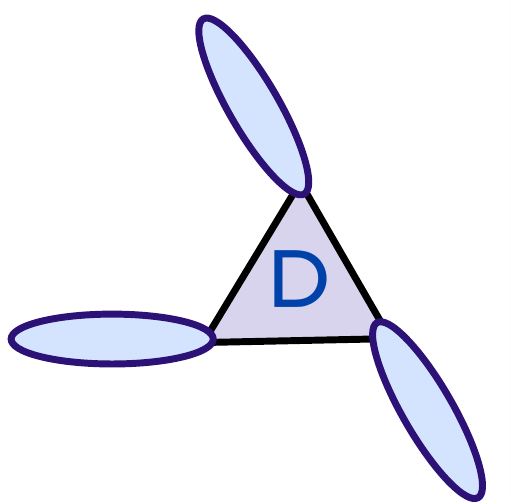,width=0.7in,clip=}}~.
\ee
Due to the defect triangle, the GS of this 6-site cluster is not the tensor product of the three singlets on the outer bonds, but it includes corrections from longer-range VBs. These corrections renormalize the GS energy from $E_0^{X_0}\!=\!-\frac{9}{4}$ to $E^{X_0}\!=\!-\frac{10}{4}$, and so the virtual fluctuations give rise to the energy correction $V_{1\text{-dt}}^{X_0}\!=\!-\frac{1}{4}$. 
To see the effect of the embedding we consider the following five clusters:
\bea
X_1 = \!\!\!\parbox{0.9in}{\epsfig{file=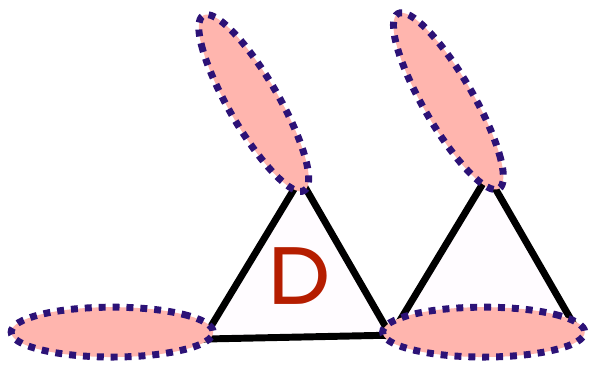,width=0.9in,clip=}},
X_2 = \!\!\!\parbox{1.15in}{\epsfig{file=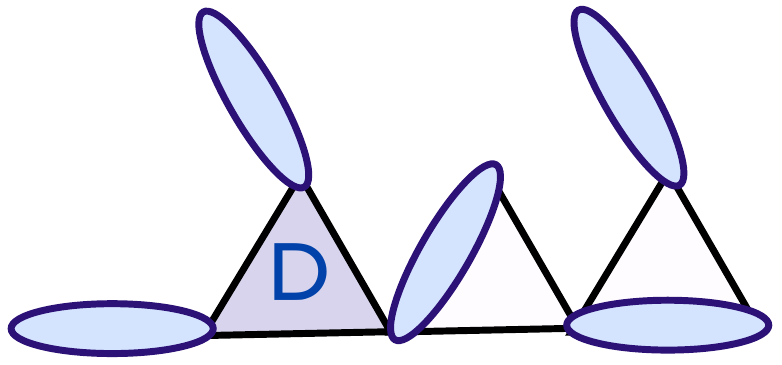,width=1.15in,clip=}},
X_3 = \!\!\!\parbox{1.15in}{\epsfig{file=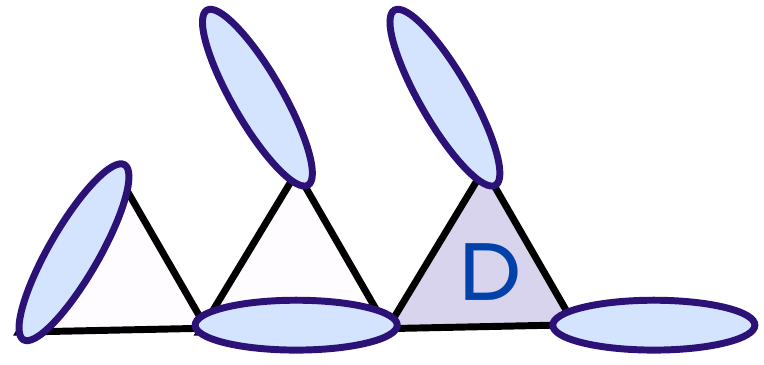,width=1.15in,clip=}},
X_4= \!\!\!\parbox{0.9in}{\epsfig{file=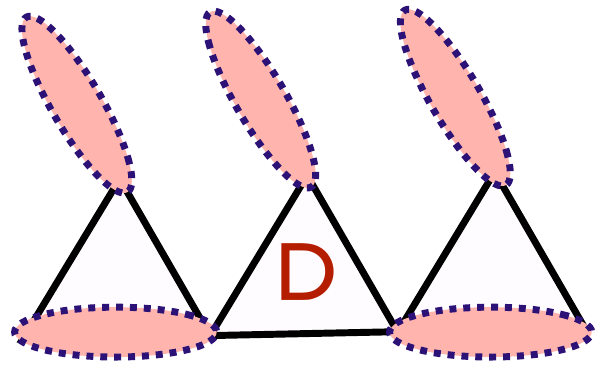,width=0.9in,clip=}},
X_5= \!\!\!\parbox{0.9in}{\epsfig{file=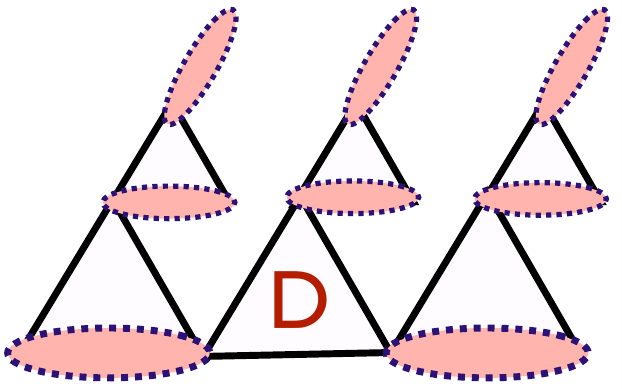,width=0.9in,clip=}}, \nonumber
\eea
which shall also be useful in the following subsection. The ground state energies of these clusters are $E^{X_1}\!=\!-3.265028$, $E^{X_2}\!=\!-4.017450$, $E^{X_3}\!=\!-4.015028$, $E^{X_4}\!=\!-4.034626$, and $E^{X_5}\!=\!-6.316796$. Subtracting the corresponding energies coming from the valence bonds,  $E_0^{X_1}\!=\!-3$, $E_0^{X_2}\!=\!E_0^{X_3}\!=\!E_0^{X_4}\!=\!-15/4$, and $E_0^{X_5}\!=\!-6$, gives the virtual corrections $V_{1\text{-dt}}^{X_1}\!=\!V_{1\text{-dt}}^{X_3}\!=\!-0.265028$~\cite{footnote}, $V_{1\text{-dt}}^{X_2}\!=\!-0.267450$, $V_{1\text{-dt}}^{X_4}\!=\!-0.284626$, and $V_{1\text{-dt}}^{X_5}\!=\!-0.316796$. %, which are all close to the value of the minimal cluster $X_0$ above. 

%%%%%%%%%%%%%%%%%%%%%%%%%%%%%%
\subsubsection{D2. Binding energy of two nearby defect triangles}
We are now ready to examine the binding energy $V_{2\text{-dt}}$ of two defect triangles. In the spirit of the numerical linked-cluster expansion~\cite{Rigol,Oitmaa}, $V_{2\text{-dt}}$ is the irreducible energy pertaining to two defect triangles, i.e. it does not include the contributions from processes involving each defect triangle alone. Consider the following two clusters
\be\label{eq:TwoDTs}
Y_1=\!\!\parbox{1.5in}{\epsfig{file=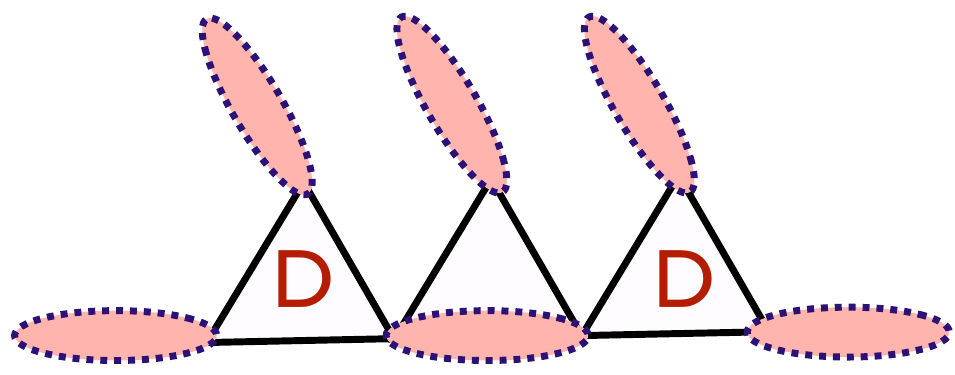,width=1.5in,clip=}},~~~~~
Y_2=\!\!\parbox{1.8in}{\epsfig{file=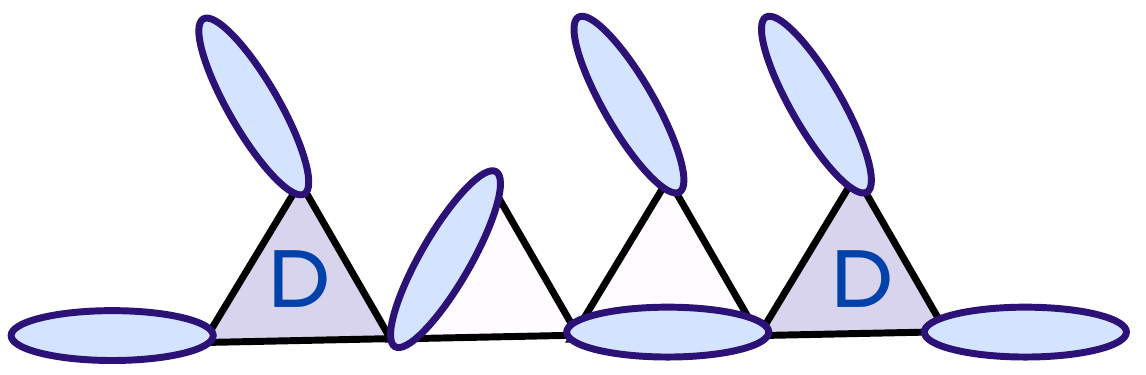,width=1.8in,clip=}},
\ee
which appear in the two NNVB states of the cluster C$_3$ of Fig.~2 of the main text. Both clusters have two defect triangles but at a different distance. 
An exact diagonalization of the Heisenberg model in these clusters gives the GS energies $E^{Y_1}\!=\!-5.037027$ and $E^{Y_2}\!=\!-5.783835$. Subtracting $E_0^{Y_1}\!=\!-9/2$ and $E_0^{Y_2}\!=\!-21/4$, respectively, gives the virtual energy corrections $E^{Y_1}\!-\!E_0^{Y_1}\!=\!-0.537027$ and $E^{Y_2}\!-\!E_0^{Y_2}\!=\!-0.533835$. To find the binding energy we need to subtract the energy corrections that come from processes that involve one defect triangle only. We have:
\bea
&&V_{2\text{-dt}}^{Y_1} = E^{Y_1}-E_0^{Y_1} - 2 V_{1\text{-dt}}^{X_1} = -0.006971,\nonumber\\
&&V_{2\text{-dt}}^{Y_2} = E^{Y_2}-E_0^{Y_2} - V_{1\text{-dt}}^{X_2} - V_{1\text{-dt}}^{X_3} = -0.001357 \,.
\eea
So we find that the binding energy in $Y_1$ is about five times larger than in $Y_2$, so we can restrict ourselves to the cases with the smallest possible distance between the two defect triangles.  
To examine the dependence on the embedding, let us consider the following clusters
\be
Y_3=\!\!\!\parbox{1.45in}{\epsfig{file=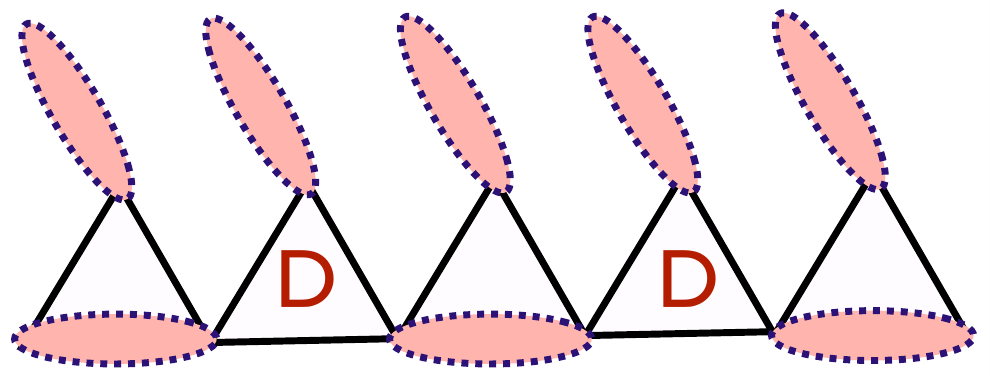,width=1.45in,clip=}},~~
Y_4=\!\!\!\parbox{1.5in}{\epsfig{file=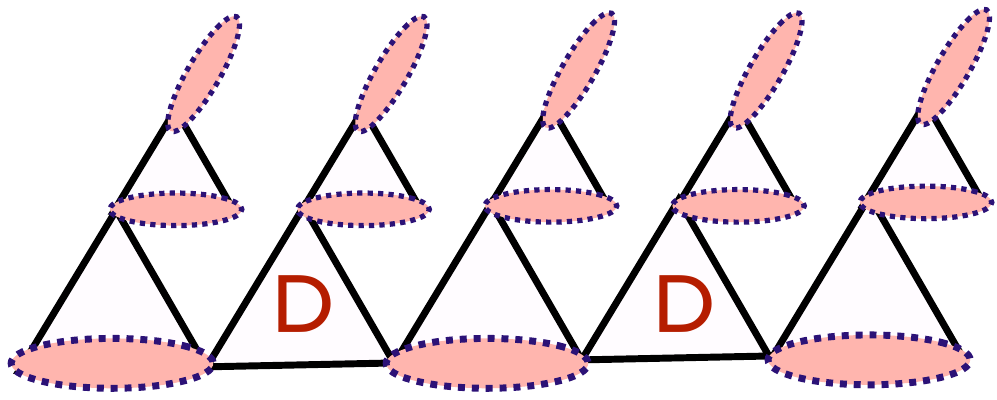,width=1.5in,clip=}}~,
\ee
which appear in the `perfect resonance' process $D$ of Fig.~2 of the main text, with $R\!=\!1$ and $R\!=\!2$ respectively. The ground state energies of these clusters are $E^{Y_3}\!=\!-6.581399$ and $E^{Y_4}\!=\!-10.394601$, the energies coming from the valence bonds are $E_0^{Y_3}\!=\!-6$ and $E_0^{Y_4}\!=\!-39/4$, and the binding energies are given by
\bea
V_{2\text{-dt}}^{Y_3} &=& E^{Y_3}-E_0^{Y_3}-2V_{1\text{-dt}}^{X_4}=-0.012147\,,\nonumber\\
V_{2\text{-dt}}^{Y_4} &=& E^{Y_4}-E_0^{Y_4}-2V_{1\text{-dt}}^{X_5}=-0.011009\,.
\eea
These values are about five times smaller than the dominant tunneling amplitudes, so altogether we can safely disregard the binding energies between two defect triangles.

%%%%%%%%%%%%%%%%%%%%%%%%%%%%%%
\subsubsection{D3. Binding energy of three defect triangles around a hexagon}
Next we turn to the binding energy of three defect triangles. The most relevant situation corresponds to having three defect triangles around a single hexagon, i.e. the `perfect hexagon' process. Consider the clusters $D(R\!=\!1)$ and $D(R\!=\!2)$ of Fig.~2 of the main text:
\be
D(R\!=\!1)=\!\!\parbox{1.in}{\epsfig{file=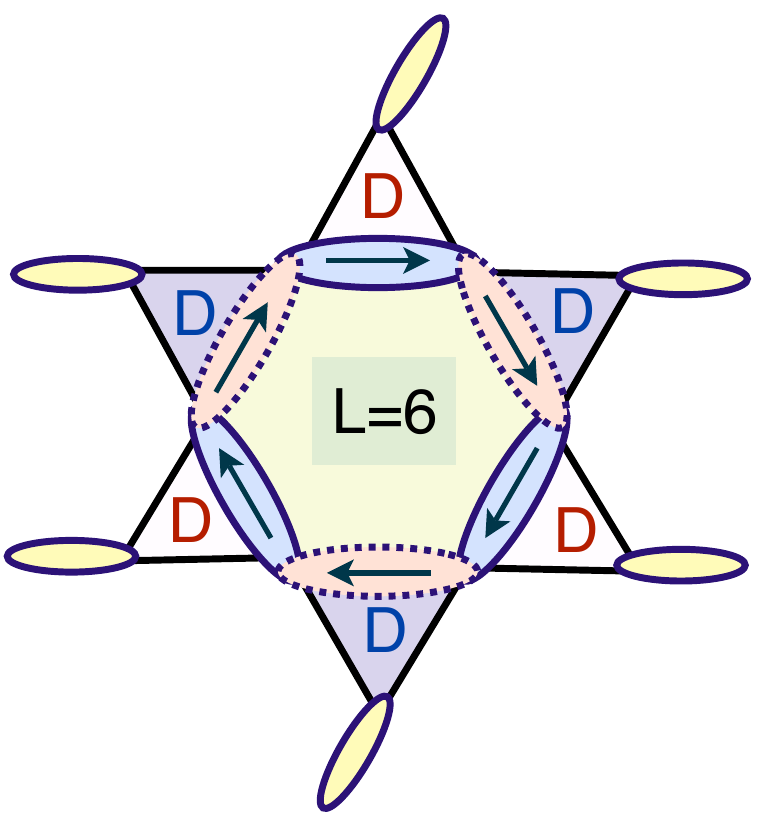,width=1.in,clip=}},~~~~~
D(R\!=\!2)=\!\!\parbox{1.in}{\epsfig{file=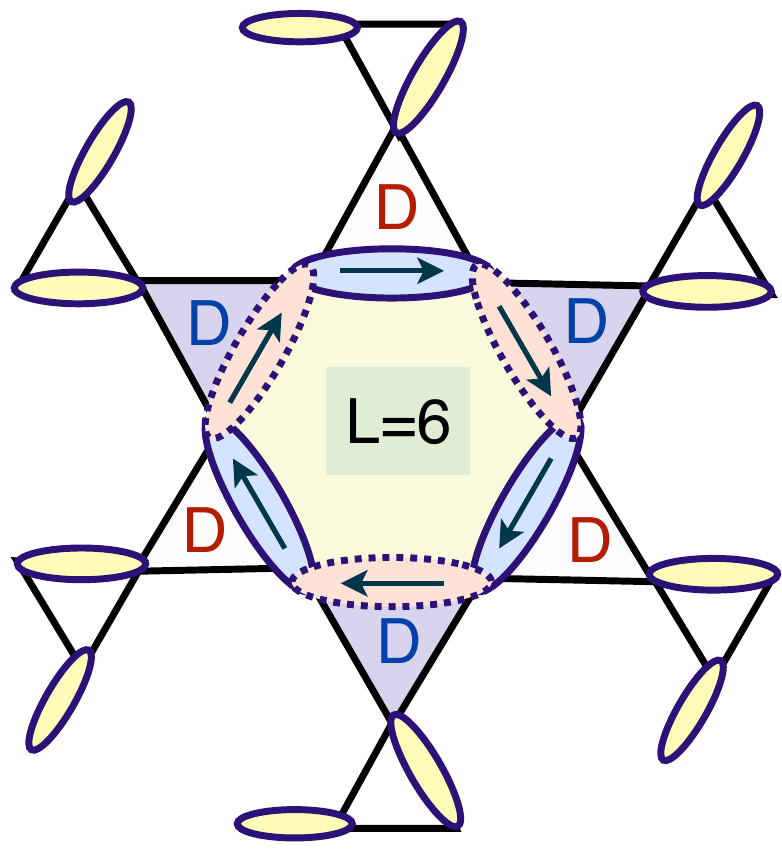,width=1.in,clip=}},
\ee
These clusters involve two NNVB states with the same potential energy $V$ (by symmetry), and to extract the later we should use the formula $V\!=\!\frac{E_+-E_-}{2}-E_0$, where $E_\pm$ are the exact tunnel-split energies. For $R\!=\!1$, $E_-^{D(R=1)}\!=\!-7.718471$, $E_+^{D(R=1)}\!=\!-7.192763$, and $E_0^{D(R=1)}\!=\!-27/4$, which gives $V^{D(R=1)}\!=\!-0.705617$. Similarly, for $R\!=\!2$, $E_-^{D(R=2)}\!=\!-12.282875$, $E_+^{D(R=2)}\!=\!-11.989591$, and $E_0^{D(R=2)}\!=\!-45/4$, which gives $V^{D(R=2)}\!=\!-0.886233$.
Now, to find the binding energy $V_{3\text{-dt}}$ of three defect triangles we must subtract the energy corrections involving a single defect triangle, and the corrections from the binding energies of two defect triangles: 
\bea
&&V_{3\text{-dt}}^{D(R=1)}=V^{D(R=1)}-3V_{1\text{-dt}}^{X_4}-3V_{2\text{-dt}}^{Y_3} = 0.184702\,,\nonumber\\
&&V_{3\text{-dt}}^{D(R=2)}=V^{D(R=2)}-3V_{1\text{-dt}}^{X_5}-3V_{2\text{-dt}}^{Y_4} = 0.097182\,.
\eea 
So the binding energy between three defect triangles around a hexagon (process `D') is the most important virtual contribution from virtual corrections to the energy. As we discuss in the main text, the fact that the binding energy is positive means that virtual fluctuations penalize the occurrence of `perfect hexagons' not only via the reduction of the tunneling amplitude $t_D$ but also via the potential terms.

%%%%%%%%%%%%%%%%%%%%%%%%%%%%%%
\section{E. Extracting the tunneling amplitude for the process $C_3$}
Finally, we come back to another issue discussed in the main text, in relation to the tunneling amplitude $t_{C_3}$ of the $C_3$ process. As we discussed there, we can extract $t_{C_3}$ from the splitting of the two low-lying singlets of the exact spectra, provided that the difference $V_r\!-\!V_b$ in the potential energies of the two NNVB states $|\phi_r\rangle$ and $|\phi_b\rangle$ involved in the tunneling process is much smaller than the splitting itself. 
The NNVB states involve two defect triangles with their distance exactly as in clusters $Y_1$ and $Y_2$ above. So we can estimate $V_r\!-\!V_b\!\approx\!V_{2\text{-dt}}^{Y_1}\!-\!V_{2\text{-dt}}^{Y_2}\!=\!0.00561$. This estimate is much smaller compared to the energy splittings found in the ED spectra of cluster C$_3$ (see Fig.~2 of the main text), meaning that we can safely take $t_{\text{ED}}$ as half of energy splitting (see main text).

\end{widetext}

%%%%%%%%%%%%%%%%%%%%%%%%%%%%%%

%%%%%%%%%%%%%%%%%%%%%%%%%%%%%%


\begin{thebibliography}{99}
\bibitem{DTDistance} Physically, this can be understood by noting that the distance between the two defect triangles involved in $|\phi_b\rangle$ is different from that in $|\phi_r\rangle$.
\bibitem{LETriplets} We remark that the presence of low-lying triplets in clusters C$_1$-C$_3$ and particularly in D is not incompatible with the NNVB picture, provided that the spin-gap remains open in the thermodynamic limit.
\bibitem{spinons} Given that dimers move by passing spinons (see discussion above), the effect of virtual fluctuations intensifies with the number of spinons involved (zero, two, four, and six in clusters A, B, C, and D, respectively).
\end{thebibliography}
\begin{thebibliography}{99}
\bibitem{footnote} Note that $V_{1\text{-dt}}^{X_1}\!=\!V_{1\text{-dt}}^{X_3}$ because the extra triangle in $X_3$ (compared to $X_1$) contains a valence bond which remains intact no matter how many times we apply the Hamiltonian.
\end{thebibliography}
\end{document}